# Parallel stability analysis of membrane lamellar structures and foam films

Nikoleta G. Ivanova and Roumen Tsekov
Department of Physical Chemistry, University of Sofia, 1164 Sofia, Bulgaria

In the frames of the DLVO theory the root mean square amplitude of capillary waves in a thin liquid film is calculated and its dependence on some important physical parameters is unveiled. Two important models are considered: films with classical interfaces and films between lipid bilayers. The performed numerical analysis demonstrates essential difference in their behavior due to the different elastic properties of the film surfaces. It is shown that the film lifetime is significantly long at some magic film radii.

Thin liquid films (TLFs) bounded by simple interfaces, insoluble monolayers and membranes are in the scientific focus from many years [1]. The interest to these nanothin structures is promoted by specific phenomena taking place in films and the large application in technology. An important aspect in the practical application is the stability of TLFs [2-4], which is decisive for many processes in flotation [5], colloid coagulation [6], food industry [7], and in other disperse systems [8]. In the literature, the TLF stability is described as a result of competition between van der Waals disjoining pressure and stabilizing effect of capillary and electrostatic forces (DLVO theory). A linear hydrodynamic theory of the kinetics of hole or black spot formation is developed [9-11] which provides calculation of the film lifetime and critical thickness. This theory accounts for many interfacial processes such as adsorption, surface and bulk diffusion of surfactants, Marangoni effect, etc. The basic restriction of the linear theory is the small parameter expansion over the ratio between the amplitude $\zeta$ of the surface corrugations and the average film thickness $h$. Since on the point of the film rupture they are of equal order such presumption is not justified. However, the linear theory results are very important for interpretation of light scattering experiments [12-15].

An interesting effect on the TLF stability is the coupling of the film drainage and the unstable modes due to non-linear TLF hydrodynamics [16, 17]. Its theoretical description requires correct calculation of the drainage rate, which depends substantially by the shape of the film interfaces. Usually, a dimple is formed [11, 18-21] and the corresponding law of thinning [22, 23] differs from the classical Reynolds expression applied first to TLF by A. Scheludko. During last decades, some attempts to take into account the influence of non-linear effects on TLF stability have been made. Some authors [24-26] have reported analytical-numerical calculations based on the Navier-Stokes equations. These theories continue usual ideas of hydrodynamic stability without considering fluctuations. Other authors [27] have accounted for the influence of the thermoconvection in a non-linear flow by means of temperature dependence of viscosity. The main non-linear effects are due to two factors: the non-linear hydrodynamic terms in the Navier-Stokes equations and the non-linear dependence of the disjoining pressure from the film thickness. In pure form, the first effect takes place at capillary-gravity waves on infinitely deep liquids where the disjoining pressure is missing. In this case, the interfacial kinematics is described by a Korteweg-de Vries equation of fifth degree [28].

The dependence of the disjoining pressure from the film thickness is a problem, which has been object of intensive investigations. Abreast of now becoming classical van der Waals and electrostatic components, a number of new interesting effects of macroscopic interaction were observed [29]. Here we can mention non-DLVO hydrophobic, hydration, protrusion, etc., forces [30]. In the case of surfactant concentration above CMC the disjoining pressure isotherm exhibits periodic behavior, which is responsible for the film stratification [31]. Similar periodic behavior but with smaller characteristic length could be expected as a consequence of the discrete molecular structure of the matter [32].

The picture in membrane multi-lamellar structures is additionally complicated by the low value of surface tension leading to large out of plane deviations of membranes. For this reason, the steric interaction between the two membranes bordered the film becomes important and leads to the so-called undulation forces [33]. This interaction depends substantially by the membrane elasticity $\kappa$, which can be conveniently measured by ellipsometry and X-ray scattering [34, 35]. Membranes are soft and easy flexible bilayer structures, which possess freedom for fluctuations into membrane plane [36]. The scale of these fluctuations grows up close to the intra-membrane phase transitions [37, 38]. The mechanical properties of membranes, even if composed by two or more monolayers of insoluble surfactants, are far from those of the latter [39, 40]. The internal behavior of bilayers is well studied by means of methods based on diffraction and scattering of light [41, 42], X-rays [43] and neutrons [44, 45], and some interesting phase transitions are observed as result of the phase state of the lyophobic parts of surfactant molecules [46, 47]. They play important role not only in the membrane transport, elastic properties and secondary quasi-crystalline arrangement of the membrane incorporations but also in the biological functions. Additional complications could arise from existence of membrane proteins, which interact each other by potential and curvature induced surface forces [21, 48, 49]. Due to the Brownian motion they migrate and create a secondary crystalline structure, which changes properties of the whole membrane [50]. The problem of rupture of TLF bounded by biomembranes is relevant to coalescence of biological cells, fusion of colloids in cells, stability of vesicles and lamellar structures and other life important processes. Both linear [51] and non-linear [52] theories were applied to the multi-lamellar membrane structure stability. A quasi-thermodynamic theory for rupture of bilayers and Newton black films was proposed [53], which explains the phenomenon by appearance of holes with a critical size.

An important feature of fluctuations in fluids is the presence of collective hydrodynamic fluctuating modes [54-57]. The local fluctuations of the TLF thickness are the major factor responsible for the film instability. Two types of modes exist in films with fluid interfaces, bending and squeezing modes, and the latter are responsible for the film rupture [13, 58]. The dynamic behavior of the squeezing modes is determined by the film disjoining pressure, surface tension and viscosity of the fluid [59-61]. The films rupture spontaneously at positive first derivative of the disjoining pressure. In the frames of the linear analysis, the pressure fluctuations can be expressed as a superposition of wave disturbances with different wave number and amplitude. It is known that asymmetric film disorders are the most destabilizing for films [60]. The effect of such surface waves is summarized in details in the paper of Narsimhan [62].

According to TLF hydrodynamics the pressure acting on the film surfaces has the form

$$p = -(\sigma/2)\Delta\zeta + (\kappa/2)\Delta\Delta\zeta - \Pi'\zeta + \delta P \tag{1}$$

The first term here represents the local capillary pressure, where σ is the surface tension. The second term accounts for elastic deformations of the film surfaces with elasticity modulus κ of the interfaces. $\Pi'$ is the first derivative in respect to $h$ of the total disjoining pressure, being sum of all existing components, and $\delta P$ represents the pressure fluctuations due to the thermal motion. In Fourier form the equation describing the thickness fluctuations reads [63]

$$(24\eta/q^2 h^3)\partial_t \zeta_q + (\sigma q^2 + \kappa q^4 - 2\Pi')\zeta_q = -2\delta P_q \tag{2}$$

where $q$ represents the wave vector modulus. One can recognize now the destabilizing role of the negative components of the disjoining pressure on the film waves. They can lead to a negative harmonic force which is an indication for divergent solutions. It is important to note the stabilizing effect of the surface tension and the film surface elasticity. It is possible to obtain from Eq. (2) the spatial spectral density of the thermal non-homogeneity of the film thickness [63]

$$C_{\zeta\zeta}(q,t) = \frac{2k_B T}{\pi R^2 (\sigma q^2 + \kappa q^4)} \frac{2\Pi' \exp[(q^2 h^3/12\eta)(2\Pi' - \sigma q^2 - \kappa q^4)t] - \sigma q^2 - \kappa q^4}{2\Pi' - \sigma q^2 - \kappa q^4} \tag{3}$$

The fundamental result (3) contains the well-known quantitative criterion for the film stability. It is easy to recognize that the thickness waves become unstable ($C_{\zeta\zeta} \to \infty$) when the inequality $2\Pi' > \sigma q^2 + \kappa q^4$ is fulfilled. The corresponding equality defines a critical wave number

$$q_{cr} = \sqrt{(\sqrt{\sigma^2 + 8\kappa\Pi'} - \sigma)/2\kappa} \tag{4}$$

below which the waves are not stable and their amplitudes grow exponentially. In the case of negligible elastic modulus κ this criterion reduces to the well-known result $q_{cr} = \sqrt{2\Pi'/\sigma}$. The other extreme case of negligible surface tension is appropriate for lamellar membrane structures and the corresponding critical wave number reads $q_{cr} = \sqrt[4]{2\Pi'/\kappa}$. The wave number of the most rapid Fourier mode is

$$q_{mq} = \sqrt{(\sqrt{\sigma^2 + 6\kappa\Pi'} - \sigma)/3\kappa} \tag{5}$$

It is easy to show that the most rapid wave is unstable, since $q_{mq} < q_{cr}$. For the two cases discussed before the corresponding values of the wave numbers of the most rapid waves are equal to $q_{mq} = \sqrt{\Pi'/\sigma}$ and $q_{mq} = \sqrt[4]{2\Pi'/3\kappa}$, respectively.

The total disjoining pressure in the DLVO theory is a sum of the electrostatic and van der Waals components. The main parameter accounting for the specific interactions in films is the first derivative of the disjoining pressure

$$\Pi' = 3K_{VW}/h^4 - 64CF\sqrt{2N_A k_B TC/\varepsilon_0 \varepsilon}\exp(-\sqrt{2C/\varepsilon_0\varepsilon N_A k_B T}Fh) \qquad (6)$$

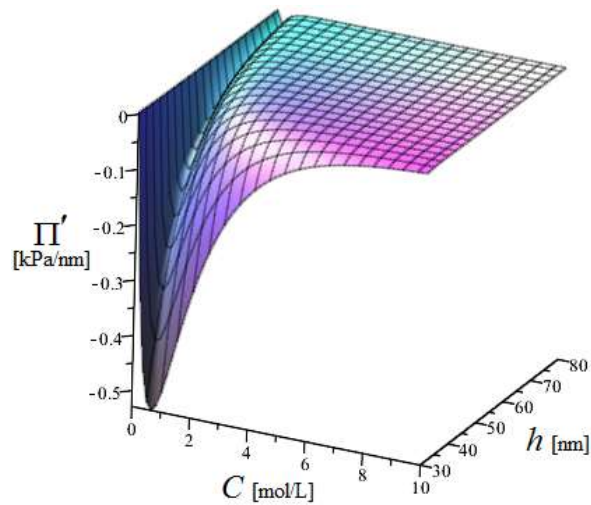

**Figure 1.** The isotherm of $\Pi'$ is presented for aqueous films

According to Fig. 1 there are two values of electrolyte concentration $C$, where the first derivative becomes zero. These lower and higher values correspond to maximum and minimum of the disjoining pressure, respectively. Unstable film possesses thickness smaller than that at the maximum at about $C$ = 10 mM, where a transition from stable to unstable film occurs. In the present paper two models are discussed with different specific constants: the Model A is a film with fluid interphases and the Model B is a film separating two lipid membranes. In both cases the fluid in the film is water and therefore the standard values of the constants are: $T$ = 300 K, $\varepsilon_0\varepsilon$ = 0.5 nF/m, $K_{VW}$ = 6.25 meV, $\eta$ = 1 mPa s, $h$ = 30 nm, $R$ = 0.1 mm. The only difference in the two models is the elastic property of the interfacial boundary. A typical value of the surface tension of the Model A is $\sigma$ = 50 mN/m, while $\kappa \equiv 0$. In this case the wave number of the most rapid mode $q_{mq}$ is about 250 mm$^{-1}$ and the critical wave number $q_{cr}$ is about 385 mm$^{-1}$. For the Model B $\sigma \equiv 0$ and $\kappa$ = 0.5 pJ, and the wave number of the most rapid mode $q_{mq}$ is about 265 mm$^{-1}$ and critical wave number $q_{cr}$ is about 345 mm$^{-1}$.

The process of film rupture occurs in two stages: formation of a hole and expansion of its area. The slow rate-determining step is the hole-formation [2, 4] and the film rupture can be defined by the appearance of the first hole in the film. It is known that small fluctuations of physical quantities are Gaussian [64]. Therefore, the probability density for a local deviation from the average thickness of the film is given by the normal distribution

$$p(\zeta,t) = \exp(-\zeta^2/2A^2)/\sqrt{2\pi A^2} \qquad (7)$$

To take into account that the film ruptures at the hole-formation, one must introduce an adsorption boundary at $\zeta = -h$. Thus, one can derive an expression for the probability density for locale thickness fluctuations in the real unstable film

$$w(\zeta,t) = [\exp(-\zeta^2/2A^2) - \exp(-(\zeta+2h)^2/2A^2)]/\sqrt{2\pi A^2} \qquad (8)$$

From Eq. (8) one can estimate the total probability for the film to be alive until the time $t$, which is equal to the probability not to have a hole in the film in the presence of adsorption boundary

$$W(t) = \int_{-h}^{\infty} w(\zeta,t)d\zeta = \mathrm{erf}(h/\sqrt{2}A) \qquad (9)$$

For unstable films $W(\infty) = 0$ and one can calculate the average lifetime of the film $\tau$ as follows

$$\tau = \int_0^{\infty} t\, d(1-W) = \int_0^{\infty} W\, dt \qquad (10)$$

This lifetime depends on the amplitude $A$ of the unstable waves and calculations are carried out numerically to discriminate the influence of different physical factors.

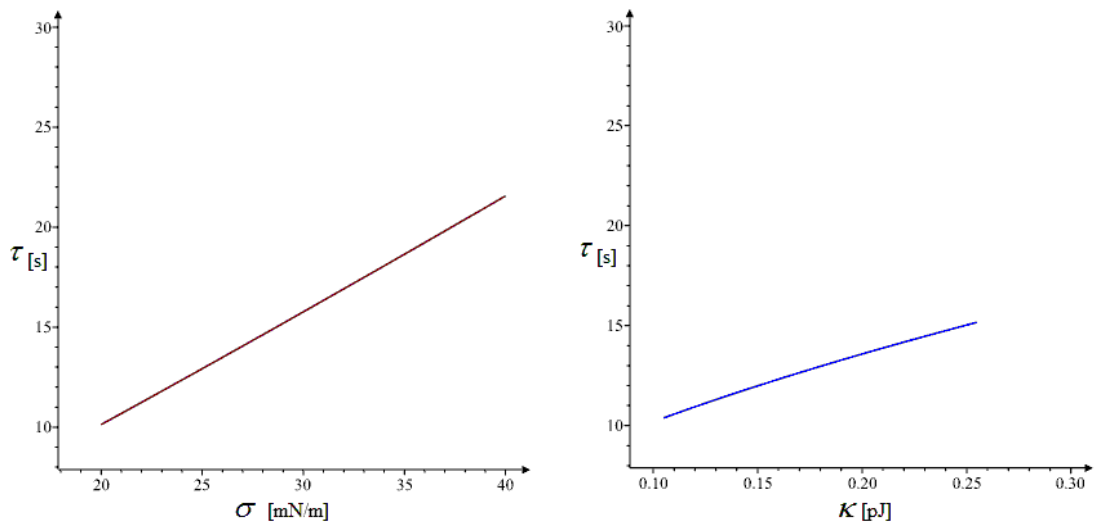

**Figure 2.** Influence of the surface tension and the elasticity modulus on the TLF lifetime

Let us consider first the influence of the elastic properties on the film lifetime (Fig. 2). The dependence of the critical value of the wave vector $q_{cr}$ for the Model A on the surface tension is the square root, while for the Model B the dependence of $q_{mq}$ on the elasticity module is the root of the square root. Hence, the lifetime of films in the Model A is stronger dependent on $\sigma$ than the dependence of $\tau$ on $\kappa$ for films in the Model B. The surface elasticity in both cases suppresses the amplitude of the fluctuation waves and thus increases the film lifetime.

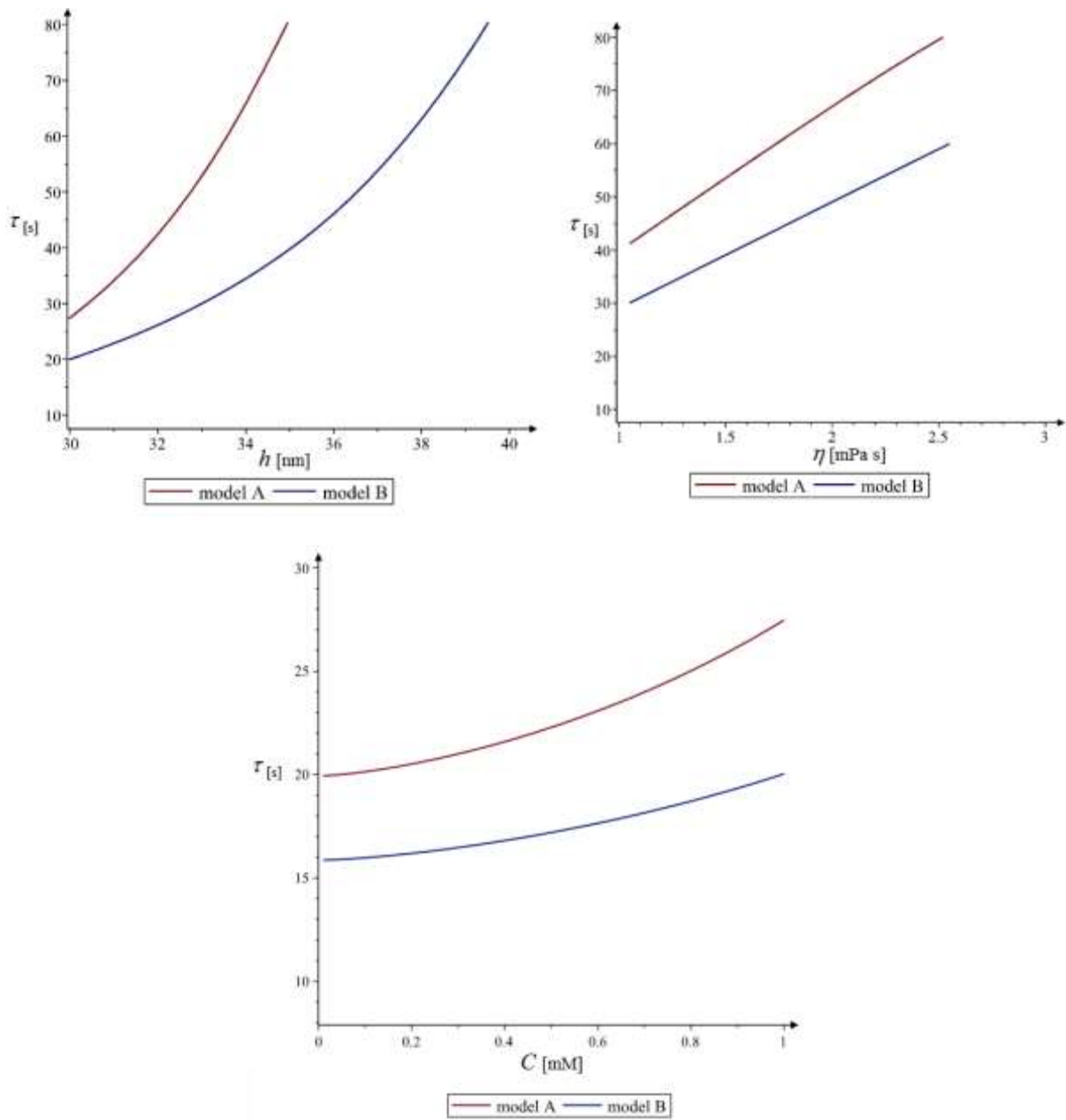

**Figure 3.** Dependences of the lifetime on film thickness, viscosity and electrolyte concentration

The dependences of the film lifetime on the film thickness, viscosity and electrolyte concentration, presented in Fig. 3 for both models, are similar. At large film thicknesses the disjoining pressure disappears and the probability of hole-formation tends to zero. This explains the observed increase of $\tau$ at higher film thickness $h$. The role of viscosity is to delay the evolution of the waves due to the increase of friction inside the film, which naturally increases the film lifetime. As is seen that the lifetime increases also with increase of the electrolyte concentration since the larger surface charge density results in higher value of the stabilizing electrostatic component of the disjoining pressure. At sufficiently large concentrations the electrostatic component must be reduced due to shortening of the electric double layer thickness. This effect should reduce also the film lifetime but it is significant for concentrations outside the considered range of $C$. The trends of the lifetime of films formed between biological membranes (the Model B) are weaker than that for the Model A.

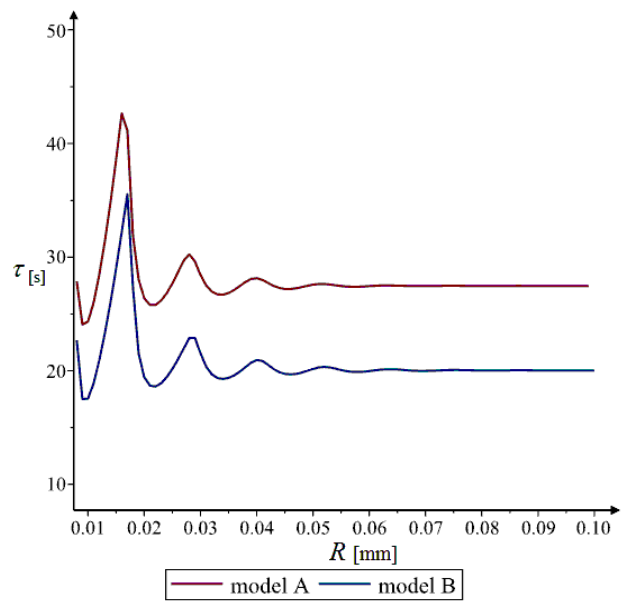

**Figure 4.** Dependence of the film lifetime on the radius of the film

Generally, the film lifetime should not depend on the film radius for non-thinning films. The behavior of $\tau(R)$ on Fig. 4 is analogous for the two considered models. There are obviously radii, at which the films are more stable due to the lack of standing waves with a wavelength equal to the most rapid one. When the most rapid wave is present in the wave spectrum, the film lifetime decreases significantly to a constant value. At larger radii this is always the case.

     In the present paper two models of thin liquid films are discussed: the Model A for films bounded by classical interfaces and the Model B for films separating lipid bilayers. As a result of the theoretical consideration differences in the behavior of the two systems are discovered, due to the different power of the elastic responses of the films. The dependences of the film lifetime on various physical parameters such as the electrolyte concentration, surface tension, interfacial elasticity, viscosity and film thickness are examined. The dependence of the lifetime on the film radius is especially interesting due to the presence of 'magic' values, corresponding to sufficiently stable films.